\documentclass[preprint]{revtex4}
\usepackage[latin9]{inputenc}
\usepackage{color}
\usepackage{amsmath}
\usepackage{amssymb}
\usepackage{graphicx}
\usepackage{esint}
\usepackage[unicode=true,pdfusetitle,
 bookmarks=true,bookmarksnumbered=false,bookmarksopen=false,
 breaklinks=false,pdfborder={0 0 1},backref=false,colorlinks=false]
 {hyperref}
\usepackage{breakurl}
\usepackage[normalem]{ulem}
\makeatletter
\@ifundefined{textcolor}{}
{%
 \definecolor{BLACK}{gray}{0}
 \definecolor{WHITE}{gray}{1}
 \definecolor{RED}{rgb}{1,0,0}
 \definecolor{GREEN}{rgb}{0,1,0}
 \definecolor{BLUE}{rgb}{0,0,1}
 \definecolor{CYAN}{cmyk}{1,0,0,0}
 \definecolor{MAGENTA}{cmyk}{0,1,0,0}
 \definecolor{YELLOW}{cmyk}{0,0,1,0}
}

\makeatother

\begin{document}

\title{Unusual eigenvalue spectrum and relaxation in the Lévy Ornstein-Uhlenbeck
process }

\author{Deepika Janakiraman }

\author{K. L. Sebastian}

\address{Department of Inorganic and Physical Chemistry, Indian Institute
of Science, Bangalore 560012, India
\\
\textbf{To appear in Physical Review E as a Rapid Communication}
}

\begin{abstract}
We consider the rates of relaxation of a particle in a harmonic well,
subject to Lévy noise characterized by its Lévy index $\mu$. Using
the propagator for this Lévy Ornstein-Uhlenbeck process (LOUP), we
show that the eigenvalue spectrum of the associated Fokker-Planck operator has the form $(n+m\mu)\nu$ where
$\nu$ is the force constant characterizing the well, and $n,m\in\mathbb{N}$.
If $\mu$ is irrational, the eigenvalues are all non-degenerate, but
rational $\mu$ can lead to degeneracy. The maximum degeneracy is
shown to be two. The left eigenfunctions of the fractional Fokker-Planck
operator are very simple while the right eigenfunctions may be obtained
from the lowest eigenfunction by a combination of two different step-up
operators. Further, we find that the acceptable eigenfunctions should have the asymptotic behavior $|x|^{-n_1+n_2\;\mu}$ as $|x| \rightarrow \infty$, with $n_1$ and $n_2$ being positive integers, though this condition alone is not enough to identify them uniquely. We also assert that the rates of relaxation
of LOUP are determined by the eigenvalues of the associated
fractional Fokker-Planck operator and do not depend on the initial
state if the moments of the initial distribution are all finite. If the initial distribution has fat tails, for which the higher moments diverge, one would have non-spectral relaxation, as pointed out by Toenjes et. al (Physical Review Letters, 110, 150602 (2013)).
\end{abstract}
\maketitle

In the recent past, diffusion processes that are anomalous have been observed in a wide variety of areas ranging from financial markets, movement of active particles in biological systems, diffusion in turbulent media etc \cite{Voit,Shepard,Bouchad}. In general, the relaxation
of a system undergoing normal diffusion is multi-exponential. The exponents are
determined by the eigenvalues of the Fokker-Planck (FP) operator associated
with the diffusion. The timescales involved are thus an intrinsic property
of the system, independent of the initial condition. For normal diffusion
of a particle in a potential $V(x)$, the associated FP operator is non-Hermitian
\cite{Risken}. However, one can make a similarity transformation to get
a Hermitian operator, analogous to the Hamiltonian operator for a
quantum system. This shows that the eigenvalues of the operator are
all real and further, by imposing physically motivated boundary conditions
(vanishing at infinity and square integrability), one finds the eigenvalue
spectrum of the operator. This spectrum determines the relaxation
characteristics of the system completely. It has been recently argued
by Toenjes \textit{et al.} \cite{nonspectral-relaxation-Sokolv} that
this traditional wisdom may not hold in general. They suggested \cite{nonspectral-relaxation-Sokolv}
that ``initial distributions which are not mapped to square integrable
functions by the similarity transformation, cannot be expanded in
terms of the eigenfunctions of the corresponding Hamiltonian operator
and will therefore relax at rates that may not be given by the Hermitian
spectrum''. This has been referred to as non-spectral relaxation.
Further, it was also suggested that ``the smallest non-spectral rate
can be smaller than the smallest spectral relaxation rate and thus,
it will dominate the relaxation behavior over the whole time range''.
This has been argued \cite{nonspectral-relaxation-Sokolv} to happen
even for the simplest of processes, viz. the Ornstein-Uhlenbeck process
(OUP).

In the recent past, a number of investigations have focussed on
processes driven by Lévy noise. Several physical problems in which they appear have been discussed in excellent reviews  \cite{Bouchad,Fogedby,Encyclopedia}. The result
of such driving is anomalous diffusion, having the displacement scaling
like $(time)^{1/\mu}$ with $0<\mu\leq2$. The process is governed
by a fractional Fokker-Planck equation \cite{metzlerthe2000,MBK_EuroPL,hanggi_PRE}, which is much more difficult
to analyze. The properties of such operators are not well understood,
as there are very few results on them. Laskin \cite{LaskinFractionalQM}
introduced a generalization of quantum mechanics referred to as fractional
quantum mechanics which has similar operators and the eigenvalue spectra
and eigenfunctions of some operators have been investigated \cite{DongJMP2007,DongJMP2008}.
Toenjes \textit{et al.} \cite{nonspectral-relaxation-Sokolv} discuss
the case of a particle in a harmonic well, subject to Lévy noise which is governed by the equation
\[
\frac{dx}{dt}=-\frac{V'(x)}{m\gamma}+\eta(t),
\]
where $V(x)$ is the potential that the particle is subjected to,
and $m$ and $\gamma$ are the mass and the friction coefficient, respectively. $\eta(t)$
is the Lévy noise, best described by its characteristic functional
$\left\langle e^{i\int_{0}^{T}dt\;\eta(t)p(t)}\right\rangle $ which
is equal to $e^{-D\int_{0}^{T}dt\;|p(t)|^{\mu}}$ \cite{chechkin_numerical,DeepikaPRE2013,Eliazar}. The resultant Lévy Ornstein-Uhlenbeck process (LOUP) is one of the
simplest of such processes. Though LOUP  appears to be quite simple, it has found applications in econophysics and statistics, to analyze
stochastic volatility models for financial assets \cite{Shepard,Voit}.
Also, the underdamped limit of LOUP has found application
in biology in describing the anomalous dynamics of cell migration \cite{Schwab}, and stochastic models for active particles \cite{Ebeling}. LOUP is governed by the corresponding fractional
Fokker-Planck equation, given by \cite{metzlerthe2000,chechkinintroduction2008}
\begin{equation}
\frac{\partial P(x,t)}{\partial t}=\left\{ -D\left(-\frac{\partial^{2}}{\partial x^{2}}\right)^{\frac{\mu}{2}}+\frac{\partial}{\partial x}\frac{V'(x)}{m\gamma}\right\} P(x,t).\label{Pot FFPE}
\end{equation}
The usual Smoluchowski equation is a special case of this and is
obtained when one puts $\mu=2$. Taking the potential to be of the
form $V(x)=\lambda x^{2}/2$, and changing over to a new variable
$D^{1/\mu}\; x\rightarrow x$ and putting $\lambda/(m\gamma)=\nu$,
we can write this as
\begin{equation}
\frac{\partial P(x,t)}{\partial t}=-\hat{H}_{\mu}P(x,t),\label{eq:LOUPEquation}
\end{equation}
with
\begin{equation}
\hat{H}_{\mu}=-\left\{ -\left(-\frac{\partial^{2}}{\partial x^{2}}\right)^{\frac{\mu}{2}}+\nu\frac{\partial}{\partial x}x\right\} .\label{non-Hermitian-Operator}
\end{equation}
The relaxation of the system governed by equations (\ref{eq:LOUPEquation})
and (\ref{non-Hermitian-Operator}) was studied by Toenjes \textit{et.
al} \cite{nonspectral-relaxation-Sokolv}. We briefly summarize their
main arguments. The relaxation is governed by the eigenvalues $\lambda$
of the operator determined by $\hat{H}_{\mu}\psi=\lambda\psi$, if
the initial state satisfies the acceptability conditions (see below).
Interestingly, one can find solutions of $\hat{H}_{\mu}\psi=\lambda\psi$
for any $\lambda$, but when one imposes acceptability conditions
on the $x$ dependence of $\psi$, only certain discrete values $\lambda_{n}$
and the associated eigenfunctions $\psi_{n}$ are allowed. In particular,
when $\mu=2$, the acceptability condition imposed is $\psi(x)e^{\nu x^{2}/4}\rightarrow0$
as $x\rightarrow\pm\infty$, which leads to $\lambda_{n}=\nu n$,
with $n\,\epsilon\,\mathbb{N}$. According to them, the relaxation
of the system, from any initial state $P_{0}(x_{i})$, which can be
expanded in terms of these eigenfunctions as
\begin{equation}
P_{0}(x_{i})=\sum_{n}c_{n}\psi_{n}(x_{i})e^{\nu x_{i}^{2}/4},\label{eq:expansion}
\end{equation}
is determined by these eigenvalues. In the above, $c_{n}$ are the
expansion coefficients. This means that the slowest relaxation would
correspond to $\nu$ (i.e., $n=1$). According to the paper, if the
initial condition cannot be expanded as in Eq. (\ref{eq:expansion}),
then the relaxation can be much slower.

On going over to Fourier space, with ${\cal P}(p,t)=\int dx\; P(x,t)e^{ipx}$,
the Eq. (\ref{eq:LOUPEquation}) becomes
\begin{equation}
\frac{\partial{\cal P}(p,t)}{\partial t}=\hat{{\cal H}}_{\mu}\;{\cal P}(p,t),\label{eq:Diffusion in momentum space}
\end{equation}
with
\begin{equation}
\hat{{\cal H}}_{\mu}=|p|^{\mu}+\nu p\frac{\partial}{\partial p}.\label{eq:LFP-operator-momentum}
\end{equation}
In the case with $\mu\neq2$, Toenjes \textit{et. al.} \cite{nonspectral-relaxation-Sokolv}
perform a similarity transformation given by $p=|\kappa|^{2/\mu}sign(\kappa)$, under which
the operator $\hat{{\cal H}}_{\mu}$ gets transformed to $\kappa^{2}+\frac{\mu\nu}{2}\kappa\frac{\partial}{\partial\kappa}$.
As this operator is similar to the one for OUP, they impose boundary
conditions appropriate for OUP and get the eigenvalues $n\mu\nu/2$,
with $n\in\mathbb{N}$, similar to that of OUP. For cases where the initial distribution
has a long tail behaving like $|x|^{-\alpha-1}$, Toenjes \textit{et. al.} claim to solve the time
evolution exactly and find that it contains time scales determined not
by $n\mu\nu/2,$ but by the numbers $\nu(n+m\mu+l\alpha)$, where
$\alpha$ is determined by the characteristics of the initial distribution
and $l,m,n\;\epsilon\;\mathbb{N}$. This does not coincide with the eigenvalue
spectrum of the operator found by the similarity transformation and
hence, the authors argue that non-spectral relaxation is the rule
rather than the exception for such processes. Note that according to their paper, the
long term relaxation is not necessarily determined by the lowest non-zero eigenvalue
of the operator $\hat{H}_{\mu}$.

In this letter, we point out that the Green's function for the operator
in Eq. (\ref{eq:LOUPEquation}) is enough to propagate any arbitrary
initial condition for any value of $\mu$, including $\mu=2$, which
is the usual OUP. We use the expression for the propagator that is
already known \cite{Fogedby,metzlerthe2000,Chechkin-2006,DeepikaPRE2013}
and find the exponents that are involved in the time evolution. They
are of the form $\nu(n+m\mu)$, thus showing that the eigenvalue spectrum
of the FP operator for the LOUP is characterized, in general, by
two ``quantum numbers'' $n\mbox{ and }m\in\mathbb{N}$
(and not one, as one would normally expect). If $\mu=1\mbox{ or }2$,
then the spectrum coincides with that for the OUP, but the degeneracies
are different for $\mu=1$. Also, we identify the left and right eigenfunctions
of the operator and arrive at a generalization of the expansion of
the propagator for OUP which is given in terms of the Hermite polynomials \cite{Risken}.
Further, we give operators that can be used to generate the right eigenfunctions
from the lowest possible eigenfunction, similar to the step-up operators
of quantum mechanics. We also discuss the boundary conditions that
when imposed on the solutions would lead naturally to the correct
identification of these eigenfunctions and eigenvalues.

\begin{figure}
\includegraphics{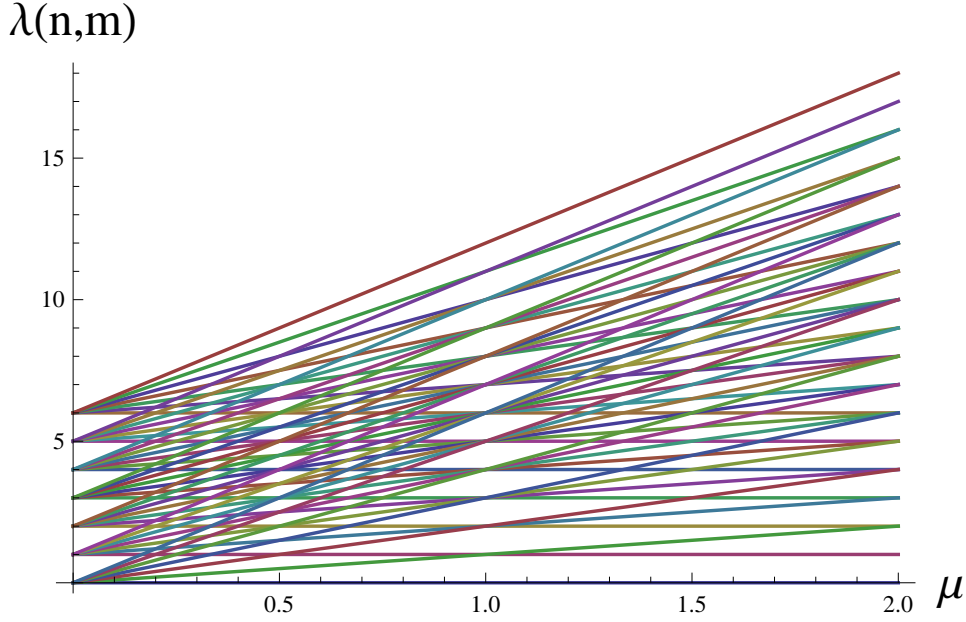} \caption{Eigenvalues $n+m\mu$ plotted as functions of $\mu$ for $\mu\;\epsilon\;[0,2]$,
for $n=0,1,2\ldots6$ and $m=0,1,2\ldots6$. The spectrum has only
non-degenerate eigenvalues if $\mu$ is irrational. Intersections
of two or more lines occur when $\mu$ is rational and can lead to
points where there is degeneracy. However, the maximum degeneracy
is only two, even though more than two lines may intersect at the
same point. At $\mu=2,$ all eigenvalues are non-degenerate, and when
$\mu=1,$ all eigenvalues except the lowest are doubly degenerate.
\label{fig. 1}}
\end{figure}

We now give our analysis of the problem. The equation (\ref{eq:Diffusion in momentum space})
can be solved by the method of characteristics, for any initial condition
${\cal P}(p,0)={\cal P}_{0}(p)$ to get the solution at a final time
$T$ as
\begin{equation}
{\cal P}(p,T)={\cal P}_{0}(pe^{-\nu T})e^{-|p|^{\mu}(1-e^{-\mu\nu T})/(\mu\nu)}.\label{eq:LOUPin momentum space}
\end{equation}
Writing ${\cal P}{}_{0}(p)=\int dx_{i}\;e^{ipx_{i}}P_{0}(x_{i})$, we
can express the position space probability distribution at the final
time $T$ as
\begin{equation}
P(x_{f},T)=\int dx_{i}\; G(x_{f},T|x_{i},0)\;P_{0}(x_{i}) \label{time-propagation}
\end{equation}
with
\begin{align}
G(x_{f},T|x_{i},0) & =\int\frac{dp}{2\pi}\; e^{-|p|^{\mu}(1-e^{-\mu\nu T})/(\mu\nu)+ip(x_{f}-x_{i}e^{-\nu T})} \label{propagator_FT}\\
= & \left(\frac{(\mu\nu)^{1/\mu}}{(1-e^{-\mu\nu T})^{1/\mu}}\right)L_{\mu}\left(\frac{(\mu\nu)^{1/\mu}\left(x_{f}-x_{i}e^{-\nu T}\right)}{\left(1-e^{-\mu\nu T}\right)^{1/\mu}}\right),\label{propagator}
\end{align}
where $L_{\mu}(x)$ is the Lévy stable distribution defined by
\begin{equation}
L_{\mu}(x)=\frac{1}{2\pi}\int^{\infty}_{-\infty} dp\; e^{ipx-\left|p\right|^{\mu}}.\label{Levy-Stable-Definition}
\end{equation}
If $\mu>1$, this may be evaluated as the series
\begin{equation}
L_{\mu}(x)=\frac{1}{\pi\mu}\sum_{n=0}^{\infty}(-1)^{n}x^{2n}\frac{\Gamma(\frac{2n+1}{\mu})}{\Gamma(2n+1)}.\label{L-expansion}
\end{equation}
We have recently developed a path integral approach to Lévy flights
which leads to exactly this result \cite{DeepikaPRE2013}. The above
analysis shows that irrespective of what the initial distribution
is, the time development of the system is determined only by the propagator. Note
that this conclusion is valid for all values of $\mu$  and is therefore
applicable to the usual Brownian motion too. It is possible to expand
the propagator $G(x_{f},T|x_{i},0)$ in terms of the left eigenfunctions
$\tilde{\psi}_{n}(x)$ and the right eigenfunctions $\psi_{n}(x)$
of the operator $\hat{H}_{\mu}$ as
\begin{equation}
G(x_{f},T|x_{i},0)=\left\langle x_{f}\left|e^{-T\hat{H}_{\mu}}\right|x_{i}\right\rangle =\sum_{n}\tilde{\psi}_{n}(x_{i})\psi_{n}(x_{f})e^{-\lambda_{n}T}.\label{eq:eigenfunction-expansion}
\end{equation}
As $\hat{H_{\mu}}$ is not a Hermitian operator, the eigenfunctions
are not necessarily orthogonal. It is obvious that if we can expand
the right hand side of Eq. (\ref{propagator_FT}) as a series in exponentials
involving $T$ (note that there are two exponentials involving $T$ leading to a double summation), then we would be able to find the eigenvalues and
eigenfunctions of the operator $\hat{H}_{\mu}$.
\begin{equation}
G(x_{f},T|x_{i},0)=\sum_{n,m=0}^{\infty}\frac{(-x_i)^n}{\Gamma(n+1) \;\Gamma(m+1)}\; \psi_{n,m}(x_{f})\;e^{-(n+m\mu)\nu T},\label{eq:spectral-expansion}
\end{equation}
where
\begin{equation}
\label{eq:righteigenfunction_FT}
\psi_{n,m}(x_{f})=\frac{1}{2\pi}\int^{\infty}_{-\infty}dp\;e^{-\frac{|p|^{\mu}}{\mu\nu}}\;(ip)^n\;|p|^{m\mu}\;e^{ipx_f}.
\end{equation}
We can now expand $e^{ipx_f}$ as a series and perform the integral over $p$ to obtain the right eigenfunction in the position space as
\begin{equation}
\, \psi_{n,m}(x_{f})=\frac{1}{\mu\pi}\sum_{k=0}^{\infty}\frac{(-1)^{\frac{(n+[n])}{2}+k}}{([n]+2k)!}\;\Gamma\left(\frac{n+[n]+2k+1}{\mu}+m\right)x_{f}^{(2k+[n])},\;\label{eq:righteigenfunction}
\end{equation}
where $[n]=n$ modulo $2$. Note that the above expansion is convergent for all $x_f$ only if  $\mu >1$. From Eq. (\ref{eq:spectral-expansion}),
it is clear that the eigenvalues of $\hat{H}_{\mu}(x)$ have the form
$(n+m\mu)\nu$ with $n$ and $m$ belonging to $\mathbb{N}.$ Thus
we have a very interesting situation that if $\mu$ is irrational,
the eigenvalues are characterized by two `quantum' numbers $n$ and
$m$, unlike the usual situation where there is only one quantum number.
Further, for such values of $\mu,$ the left eigenfunction is simply
$\tilde{\psi}_n=(-x_{i})^{n}/\Gamma(n+1)$ while the right eigenfunction is given by
$\psi_{n,m}(x_{f})$. If $\mu$ is a rational number, the left eigenfunction can become a little bit more complex. This point will be elaborated in the later part of this paper. Eq. (\ref{eq:spectral-expansion})
is the generalization of the classic expansion of the propagator for
Ornstein-Uhlenbeck process \cite{Risken}, to the LOUP. The eigenvalue
spectrum for LOUP is shown in Fig. \ref{fig. 1} as a function of
$\mu$. We also note that to use the expansion (\ref{eq:spectral-expansion}) and to calculate
$P(x_f,T)$ using Eq. (\ref{time-propagation}), it is necessary that {\it all the moments of the initial
distribution $P(x_i)$ should exist}.  If one chooses an initial distribution for which the moments diverges (for example $P(x_i)=\frac{1}{w_0}L_\alpha(x_i/w_0)$) then one cannot use Eq. (\ref{eq:spectral-expansion}) and the
relaxation would contain other time scales \footnote{We thank the authors of reference [2] for
pointing this out.} leading to non-spectral relaxation as pointed out in
\cite{nonspectral-relaxation-Sokolv}. However, if the initial distribution is a truncated L\'{e}vy distribution, then the relaxation is purely spectral.

It is possible to express the eigenfunctions in an interesting way.
If one allows $T\rightarrow\infty,$ $G(x,T|x_{i},0)$ approaches
the steady equilibrium state and becomes
\[
\psi_{0,0}(x)=(\mu\nu)^{1/\mu}L_{\mu}\left((\mu\nu)^{1/\mu}\; x\right).
\]
This is the lowest eigenfunction of the operator $\hat{H}_{\mu}$,
having the eigenvalue zero. It is easy to prove the following commutation
relations:
\begin{equation}
\left[\frac{\partial}{\partial x},\hat{H}_{\mu}\right]=-\nu\frac{\partial}{\partial x},\label{eq:commutation-relation-1}
\end{equation}
and
\begin{equation}
\left[\left(-\frac{\partial^{2}}{\partial x^{2}}\right)^{\mu/2},\hat{H}_{\mu}\right]=-\mu\nu\left(-\frac{\partial^{2}}{\partial x^{2}}\right)^{\mu/2}.\label{eq:commutation-relation-2}
\end{equation}
The above imply that if $\psi$ is an eigenfunction of $\hat{H}_{\mu}$
with an eigenvalue $\varepsilon$, then $\frac{\partial\psi}{\partial x}$
and $\left(-\frac{\partial^{2}}{\partial x^{2}}\right)^{\mu/2}\psi$,
too are eigenfunctions with eigenvalues $\varepsilon+\nu$ and $\varepsilon+\mu\nu$,
respectively. It follows that the right eigenfunction $\psi_{n,m}(x)\propto\left(-\frac{\partial^{2}}{\partial x^{2}}\right)^{\mu m/2}\frac{\partial^{n}}{\partial x^{n}}\psi_{0,0}(x)$.
The asymptotic behavior as $x\rightarrow\pm\infty$
is
\begin{equation}
\psi_{n,m}(x)\sim\frac{1}{|x|^{n+1+\mu}}\mbox{ when }m=0\label{right_eigen_asym_m_0}
\end{equation}
and
\begin{equation}
\psi_{n,m}(x)\sim\frac{1}{|x|^{n+1+m\mu}}\mbox{ when }m\neq0\label{right_eigen_asym_m_non_0}
\end{equation}
However, one can find eigenfunctions other than these, as shown
below. This is easily done in the momentum space.

The eigenvalue equation in the momentum space
\begin{equation}
\left(|p|^{\mu}+\nu p\frac{\partial}{\partial p}\right)\overline{\psi}_{\lambda}(p)=\nu\lambda\overline{\psi}_{\lambda}(p),\label{eq:LFP-momentum-eigenvalue-equation}
\end{equation}
has the solution
\begin{equation}
\overline{\psi}_{\lambda}(p)=e^{-\left|p\right|^{\mu}/(\mu\nu)}p^{\rho}\left|p\right|^{\sigma},\label{eq:momentum-eigen-function}
\end{equation}
for any real positive numbers $\rho$ and $\sigma$ such that $\lambda=\rho+\sigma$. Its position space representation is given by
\begin{equation}
\label{eq:momentum-eigen-function_FT}
\psi_{\lambda}(x)=\frac{1}{2\pi}\int_{-\infty}^{\infty}dp\; e^{-|p|^{\mu}/(\mu\nu)+ipx}p^{\rho}\left|p\right|^{\sigma}.
\end{equation}
For an arbitrary  $\lambda=\rho+\sigma$, we may rearrange  Eq. (\ref{eq:momentum-eigen-function_FT}) to get
\begin{eqnarray}
\label{eq:phi_p}
\psi_{\lambda}(x) & = &\frac{1}{2\pi}\int_{-\infty}^{\infty}dp\; e^{-|p|^{\mu}/(\mu\nu)+ipx}\left|p\right|^{\lambda}(\mbox{sign}p)^{\rho}\nonumber\\
&&\nonumber\\
&=&\frac{(1+e^{i\pi\rho})}{2\pi}\; C_{\lambda}(x)+\frac{(1-e^{i\pi\rho})}{2\pi}\; S_{\lambda}(x),
\end{eqnarray}
with
\begin{equation}
C_{\lambda}(x)=\int_{0}^{\infty}dp\; e^{-p^{\mu}/(\mu\nu)}p^{\lambda}\cos(px)\label{C_int}
\end{equation}
\begin{equation}
S_{\lambda}(x)=\int_{0}^{\infty}dp\; e^{-p^{\mu}/(\mu\nu)}p^{\lambda}\sin(px).\label{S_int}
\end{equation}
From the momentum space eigenfunction in Eq. (\ref{eq:momentum-eigen-function}), it appears that a continuous infinity of $\lambda$ will satisfy the differential equation in Eq. (\ref{eq:LFP-momentum-eigenvalue-equation}). However, we find that $\lambda=n+m\mu$ are the only eigenvalues we find in the spectrum. The question that we ask is what is the sanctity of these $\lambda=n+m\mu$ and why are the other values of $\lambda$ unacceptable?
In order to answer this question, we now look at the asymptotic behavior
of the two linearly independent solutions in the position space for the eigenvalue equation Eq. (\ref{eq:LFP-momentum-eigenvalue-equation}), viz. $C_{\lambda}(x)$ and $S_{\lambda}(x)$ which have the same eigenvalue $\nu \lambda$. For this we use the identity
\[
\cos(px)=\frac{1}{2i}\int_{c-i\infty}^{c+i\infty}ds\;\frac{\Gamma(s)}{\Gamma(\frac{1-s}{2})\Gamma(\frac{1+s}{2})}(px)^{-s},
\]
where $c>0$ (note the the right hand side of the equation is a Mellin-Barnes
integral) in the expression for $C_{\lambda}(x)$ and then perform
the integral over $p$, to get
\begin{equation}
C_{\lambda}(x)=\frac{(\mu\nu)^{\frac{\lambda+1}{\mu}}}{2i\mu}\int_{c-i\infty}^{c+i\infty}ds\;\frac{\Gamma(s)\Gamma\left(\frac{1+\lambda-s}{\mu}\right)}{\Gamma(\frac{1+s}{2})\Gamma(\frac{1-s}{2})}\left((\mu\nu)^{1/\mu}x\right)^{-s}.\label{exp_C}
\end{equation}
If one closes the contour on the left hand side using a semi-circle
of radius $R\rightarrow\infty$, one gets a series expansion in terms
of $x^{n}$, while if one closes the contour on the right hand side,
one gets an asymptotic expansion in inverse powers of $x$. This asymptotic
expansion is
\begin{equation}
C_{\lambda}(x)=\frac{(\mu\nu)^{\frac{\lambda+1}{\mu}}}{\mu}\sum_{k=0}^{\infty}(-1)^{k}\;\frac{\Gamma(\lambda+k\mu+1)}{\Gamma(k+1)}\cos\left(\frac{\pi}{2}(\lambda+k\mu+1)\right)\;\left((\mu\nu)^{1/\mu}x\right)^{-(\lambda+k\mu+1)}.\label{C_x}
\end{equation}
$S_{\lambda}(x)$ can be evaluated in an exactly similar fashion to
be
\begin{equation}
S_{\lambda}(x)=\frac{(\mu\nu)^{\frac{\lambda+1}{\mu}}}{\mu}\sum_{k=0}^{\infty}(-1)^{k}\;\frac{\Gamma(\lambda+k\mu+1)}{\Gamma(k+1)}\sin\left(\frac{\pi}{2}(\lambda+k\mu+1)\right)\;\left((\mu\nu)^{1/\mu}x\right)^{-(\lambda+k\mu+1)}.\label{S_x}
\end{equation}
From these expressions, we can see that unless $\lambda=n+m\mu$, the functions will not have the correct asymptotic behaviour as prescribed by Eq. (\ref{right_eigen_asym_m_0}) and Eq. (\ref{right_eigen_asym_m_non_0}). Interestingly even imposing this condition is
not enough to restrict eigenfunctions and eigenvalues to those
found from the Green's function, because for any $\lambda$ we still seem to be having
two solutions, viz. $C_{\lambda}(x)$ and $S_{\lambda}(x)$. We now
discuss how to identify the correct eigenfunctions which appear in
the expansion of Eq. (\ref{eq:spectral-expansion}).

For any value of $\lambda$, we can write down the different possible
ways in which the eigenvalue can have the form $n+m\mu$. We now consider
two separate possibilities:
\begin{enumerate}
\item {\it $\mu$ is irrational:} For any $\lambda=n+m\mu$, the values $(n,m)$ are unique.
Further, from the Eq. (\ref{C_int}) and (\ref{S_int}), it is clear that if $n$ is even, then $C_n(x)$ behaves
like $|x|^{-n-\mu-1}$, and $S_n(x)\sim|x|^{-n-1}$ and hence, only $C_n(x)$
is acceptable. The function $C_{n+m\mu}(x)$ is obtained from $C_n(x)$ by the application of $\left ( -\frac{\partial^2}{\partial x^2} \right )^{\mu/2}$ $m$ times, an operation which does not change its symmetry. Therefore, $C_{n+m\mu}$ becomes the acceptable eigenfunction for $\lambda=n+m\mu$ where $n$ is even, as $C_n(x)$ itself is acceptable. $S_{n+m\mu}(x)$ is an unacceptable solution when $n$ is even since $S_n(x)$ itself is not. One can make a similar argument when $n$ is odd.


\item {\it $\mu$ is rational:} $\mu$ can be written as $p/q$, where $p$
and $q$ are integers having no common factors. Then one can have
degeneracy if $m$ is an integral multiple of $q$ equal to $kq$.
The associated eigenvalue is $(n+kp)\nu$. Then all the states having
quantum numbers $(n+kp,0)$, $(n+(k-1)p,q),$ $(n+(k-2)p,2q)\ldots$$(n,kq)$
will have the same eigenvalue and hence, one expects degeneracy. However,
if $n$ and $p$ even, then all these possibilities lead to the even
eigenfunction, viz. $C_{n+kp}(x)$ and the level will be non-degenerate.  On the other hand, if either $n$ or $p$ is odd or both are odd, then one of the states
will be $S_{n+kp}(x)$ and the other will be $C_{n+kp}(x)$. Thus one would have
degeneracy in this case, and the degeneracy would be two.
\end{enumerate}

Another interesting observation about the degeneracies of states has emerged out this analysis. If we consider a particular state say $\lambda=4$ for $\mu=1$ in Fig. \ref{fig. 1}, we see that five lines intersect at this point, seeming to suggest that the degeneracy of this state is $5$. However, from the previous arguments the degeneracy of this state can be found to be $2$. Of the five states which intersect at $\lambda=4$, three of them become exactly identical to $C_4(x)$ and two of them  to $S_4(x)$ and therefore, are not independent states. When the right eigenfunctions become identical, the corresponding left eigenfunctions, $(-x_i)^n/\Gamma(n+1)$, will add up to give a single left eigenfunction. This sudden reduction in the number of eigenstates as one goes from $\mu=1-\epsilon$ to $\mu=1$ to  where $\epsilon$ is infinitesimally small is not a problem as they are not constrained to be orthogonal, as is the case in quantum mechanics.   This can also be treated as an evidence for the lack of a similarity transformation for the LOUP operator converting it to a Hermitian operator, as proposed by Toenjes \textit{et al.} If such a transformation existed, it seems impossible for the eigenfunctions to undergo this sudden reduction in their number.

In conclusion, we have shown that the eigenvalue spectrum of the Fokker-Planck operator
for LOUP to be of the form $n+m\mu$, characterized by the two `quantum
numbers' $n,m\in\mathbb{ N}.$ Using the spectral expansion of the propagator, we have found the
left and right eigenfunctions of the operator. For irrational values of $\mu$ the spectrum
is non-degenerate, while for rational $\mu$ there could be degeneracies with
the maximum degeneracy being two. We also find that any acceptable eigenfunction
of the operator should satisfy the condition that as $|x|\rightarrow\infty$,
the functions should behave like $|x|^{-(n_{1}+\mu\; n_{2})}$, where $n_1$ and $n_2$ are positive integers, though this condition alone is not enough uniquely identify the eigenstates. If the moments of the initial distribution are all finite, then the relaxation is governed only by these eigenvalues, while for initial distributions having long tails, one would have non-spectral relaxation in agreement with \cite{nonspectral-relaxation-Sokolv}.

We thank the authors of reference  \cite{nonspectral-relaxation-Sokolv} for their comments and a crucial clarification.  We also thank Diptiman Sen for useful comments.  The work of both the authors is supported by Department of Science and Technology, Govt. of India through the J C Bose Fellowship of K L Sebastian.

\bibliographystyle{apsrev}

\end{document}